# Catalysts for the hydrogen evolution reaction in alkaline medium: Configuring a cooperative mechanism at the Ag-Ag$_2$S-MoS$_2$ interface

Avraham Bar-Hen[a,b], Simon Hettler[c,d], Ashwin Ramasubramaniam[e], Raul Arenal[c,d,f], Ronen Bar Ziv[b,*], Maya Bar Sadan[a,*]

[a] *Department of Chemistry, Ben-Gurion University of the Negev, Beer-Sheva-8410501, Israel*

[b] Department of Chemistry, *Nuclear Research Center Negev, P.O. Box 9001, Beer-Sheva 84190, Israel*

[c] *Laboratorio de Microscopías Avanzadas, Universidad de Zaragoza, 50018 Zaragoza, Spain*

[d] *Instituto de Nanociencia y Materiales de Aragon, CSIC-U. de Zaragoza, Calle Pedro Cerbuna 12, 50009 Zaragoza, Spain*

[e]*Department of Mechanical and Industrial Engineering, University of Massachusetts, Amherst, Massachusetts 01003, United States*

[f] *ARAID Foundation, 50018 Zaragoza, Spain.*

**\* Corresponding authors.**

*E-mail addresses*: barsadan@bgu.ac.il (Bar Sadan M.), bronen@post.bgu.ac.il (Bar-Ziv R.)

**Abstract**

Designing electrocatalysts for HER in alkaline conditions to overcome the sluggish kinetics associated with the additional water dissociation step is a recognized challenge in promoting the hydrogen economy. To this end, delicately tuning the atomic-scale structure and surface composition of nanoparticles is a common strategy and, specifically, making use of hybrid structures, can produce synergistic effects that lead to highly active catalysts. Here, we present a core-shell catalyst of Ag@MoS$_2$ that shows promising results towards the hydrogen evolution reaction (HER) in both 0.5 M H$_2$SO$_4$ and 0.5

M KOH. In this hybrid structure, the $MoS_2$ shell is strained and defective, and charge transfer occurs between the conductive core and the shell, contributing to the electrocatalytic activity. The shelling process results in a large fraction of $Ag_2S$ in the cores, and adjusting the relative fractions of Ag, $Ag_2S$, and $MoS_2$ leads to improved catalytic activity and fast charge-transfer kinetics. We suggest that the enhancement of alkaline HER is associated with a cooperative effect of the interfaces, where the Ag(I) sites in $Ag_2S$ drive the water dissociation step, and the formed hydrogen subsequently recombines on the defective $MoS_2$ shell. This study demonstrates the benefits of hybrid structures as functional nanomaterials and provides a scheme to activate $MoS_2$ for HER in alkaline conditions.

1. **Keywords**: Water splitting; Core-shell; Catalytic mechanism; Electrocatalysis; 2D materials

## Introduction

Producing hydrogen as a renewable energy source by electrocatalytic water splitting via the hydrogen evolution reaction (HER) has been considered an environmentally friendly path to replace fossil fuels [1–5]. Pt-based materials are the best HER electrocatalysts known to date but are expensive and scarce. Moreover, catalysts based on Pt-group metals have limited catalytic performance in alkaline medium compared to acidic conditions, reducing their applicability towards chlor-alkali and alkaline water electrolysis, which is considered the most technologically mature system for $H_2$ production [6]. Therefore, it is highly desired to develop catalysts based on earth-abundant transition metals as an alternative [4,7,8]. Currently, most of the researchers have focused on HER in acidic medium in which the hydrogen binding energy is considered the sole activity descriptor for predicting the performance of newly developed catalysts, while the design and understanding of the mechanistic aspect of catalysts for alkaline HER are yet to be formulated [7,9,10]. Similar to the HER pathway in acidic medium, the alkaline HER also occurs via Volmer-Tafel or Volmer-Heyrovsky paths, but the hydrogen intermediate (H*) following the Volmer step is formed by the initial water dissociation step, i.e., $H_2O + e^- \rightarrow H^* + OH^-$ [6,11,12]. This electron transfer through water dissociation may introduce an additional energy barrier and govern the overall reaction rate [12].

Molybdenum disulfide ($MoS_2$), one of the common transition-metal dichalcogenides, has emerged as a promising, low-cost, precious-metal-free catalyst. It has become the focus for research on HER electrocatalysis because of the unique layered structure and the close to zero value of the free energy of H* adsorption ($\Delta G_H$), which is a crucial descriptor of the hydrogen evolution reaction (HER) [7,13,14].

However, according to theoretical calculations and experiments, the number of active sites for HER is limited, and located along the edges of MoS$_2$. At the same time, the basal planes are relatively inert [15,16]. The fewer exposed active sites and low conductivity of MoS$_2$ sheets significantly impair the charge transfer rate, thus decreasing the HER efficiency [7]. So far, various pathways were developed to tackle these challenges, such as doping or functionalizing the MoS$_2$ to improve its intrinsic conductivity [17–23], developing the synthesis of other morphologies [24], or creating structural defects [25–31]. Combining MoS$_2$ with additional materials in a core-shell structure can effectively promote active sites as the curvature makes the MoS$_2$ shell strained and defective [32–34]. Previous studies demonstrated theoretically and experimentally that the catalytic activity of a core-shell structure could be enhanced by charge transfer from the conductive core to MoS$_2$ shell [35, 36], but these were tested mainly in acidic conditions and developing catalysts for an alkaline medium is still required. In alkaline medium, the attractive properties of the MoS$_2$ edge sites for adsorption and recombination of the H*$_{ads}$ intermediate are scaled down due to the high kinetic barrier required for the coupled process of electron transfer and water dissociation (the Volmer step) [37,38]. M$^0$/M$_2$S-MoS$_2$ (M= Ag, Cu) is a system that has attracted attention due to its HER performance which is based on enhanced electron transport [36,39,40]. The origin and mechanism of the improved electrocatalytic activity of the system has not yet uncovered, particularly in alkaline medium. So far, the influence on the catalytic performance of the various parameters such as the nature of the core–shell interactions, the degree of wrapping of the core by the shell, the presence of different chemical species and the cooperative effect of interfaces is still not well understood. Moreover, many of the structures hold a geometry where the MoS$_2$ layers are orthogonal to the surface, and a relatively thick shell of MoS$_2$ is formed. This morphology correlates the accelerating of the catalytic activity with the 3D structure and porosity of the MoS$_2$ shell. It can be assumed that the core is more than a template to the MoS$_2$ shell [41–43], and Ag core offers additional benefits due to its exquisite properties for electron capture and transport [42,44].

Herein, we propose a design concept to enhance the HER activity of MoS$_2$ in acidic and alkaline electrolytes by shelling Ag cores with a monolayer of MoS$_2$. The MoS$_2$ shell produced by the process is strained and defective, potentially comprising S vacancies that are known to be favorable catalytic sites for H$_{ads}$, a necessary feature in acidic media but insufficient for alkaline HER. The Ag@MoS$_2$ hybrids contain a large fraction of Ag$_2$S, and by optimizing the relative fractions of Ag, Ag$_2$S, and MoS$_2$, we are able to improve the performance of the hybrids, as well as gain insights to the catalytic mechanism. We propose that the enhancement is associated with a cooperative effect in alkaline medium, where the Ag(I) sites in $Ag_2S$ are used for the water cleavage step and the formed hydrogen subsequently recombines on the defective $MoS_2$ shell. The metallic core improves the conductivity of the hybrid (since both $MoS_2$ and $Ag_2S$ are semiconductors) and provides charge transfer to $MoS_2$. The hybrids demonstrate a unique platform to design and construct efficient, durable, and cost-effective electrocatalysts.

## 2. Experimental

The hybrid structures were prepared by reducing Ag(I)$_{(aq)}$ using NaBH$_4$ to form Ag nanoparticles and then slowly adding (NH$_4$)$_2$MoS$_4$ (Ammonium tetrathiomolybdate, ATM) in various ratios, to final concentrations of 0.05–0.60 mg L$^{-1}$ of (NH$_4$)$_2$MoS$_4$. The suspension was stirred overnight to bind the MoS$_4^{2-}$ anions to the Ag core, centrifuged to collect the nanoparticles, and dried and annealed under an inert atmosphere in a quartz ampoule. In addition, free-standing few-layered MoS$_2$, Ag and Ag$_2$S nanoparticles were also produced as reference samples. Details of the procedures for the synthesis and the characterization techniques are provided in the Supporting Information. Working electrodes were prepared by dispersing the catalyst, carbon black, and Nafion to produce a homogeneous ink. The ink was drop-casted onto a polished glassy carbon electrode. Electrocatalytic measurements were conducted in Ar-saturated 0.5 M H$_2$SO$_4$ and 0.5 M KOH as electrolytes for HER using a three-electrode system: the active material comprised the working electrode, the counter electrode was a graphite rod, and Ag/AgCl was the reference electrode. All the potentials refer to the reversible hydrogen electrode (RHE). For more details on the electrochemical measurements, EIS and ECSA (electrochemical surface area) tests, see the Supporting Information.

## 3. Results and discussion

Hybrids prepared with low ATM concentrations (0.05 mg L$^{-1}$) showed bare Ag structures without apparent MoS$_2$ as a shell (Fig. 1a). Since the ATM anions also serve as capping ligands to stabilize the nanoparticles during the synthesis, their low concentration led to aggregation of the Ag structures. Increasing the ATM concentrations (0.20–0.60 mg L$^{-1}$) resulted in nanoparticles covered with 1–2 layers of MoS$_2$, without free-standing MoS$_2$ (Fig. 1b–d). Transmission electron microscopy (TEM) analyses, in particular energy dispersive X-ray spectroscopy (EDS) and electron energy loss spectroscopy (EELS), confirmed the existence of Ag, Mo and S (Fig. S1a and b). TEM images of the Ag@MoS$_2$ hybrids (Ag@MoS$_2$-0.20) showed a curved and defective MoS$_2$ layer, containing dislocations and grain boundaries that create abundant MoS$_2$ edge sites on the surface (Fig. S1c). SEM images of the hybrids are also provided in the supporting information (Fig. S2).

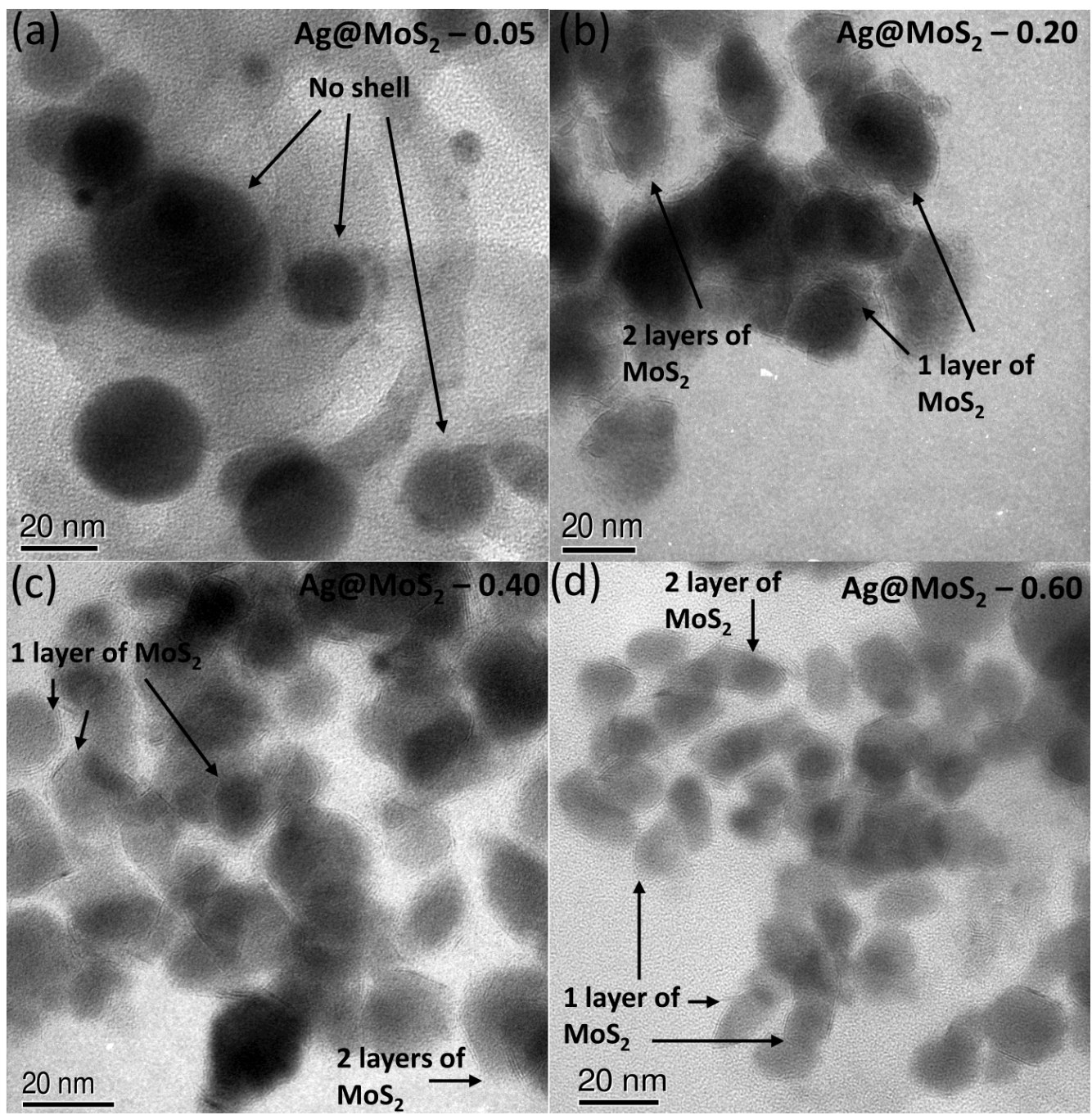

**Fig. 1.** TEM images of Ag@MoS$_2$ hybrids with various ATM concentrations (in mg L$^{-1}$). (a) Ag@MoS$_2$-0.05. (b) Ag@MoS$_2$-0.20. (c) Ag@MoS$_2$-0.40. (d) Ag@MoS$_2$-0.60.

Powder X-ray diffraction (XRD) patterns of the hybrids contain signals from 2H-MoS$_2$, Ag$_2$S and metallic Ag peaks (Fig. 2a) [40,45–47], indicating that the Ag cores are sulfidized, at least partially, to Ag$_2$S, potentially due to the release of sulfur during the decomposition of ATM to form MoS$_2$ [40]. The absence of the (002), (103), and (105) lines of the 2H-MoS$_2$ diffraction pattern is a strong evidence for single-layer MoS$_2$ [40,43]. Raman spectroscopy measurements of the Ag@MoS$_2$ hybrids showed the typical peaks of MoS$_2$ layers for all the hybrids set (Fig. 2b). The separation between the E$^1_{2g}$ and A$_{1g}$ peaks, which is generally correlated in the literature with additional layers of MoS$_2$, varied from 20 to 21 cm$^{-1}$, corresponds to one layer of MoS$_2$ [48]. Moreover, for most of the Ag@MoS$_2$ hybrids (0.20-0.60), a pronounced redshift of the Raman peaks is observed relative to bulk MoS$_2$, which is associated with strain in the MoS$_2$ layers [49]. Specifically, the shift of the E$^1_{2g}$ to lower frequencies, i.e., below 380 cm$^{-1}$, is considered evidence of S vacancies in the MoS$_2$ shell [50–52]. Ag@MoS$_2$-0.40 hybrids exhibit the lowest E$^1_{2g}$/A$_{1g}$ peak ratio (Table S1), which was correlated with a higher fraction of exposed MoS$_2$ edges, some of which are also domain boundaries caused by S deficiencies [53].

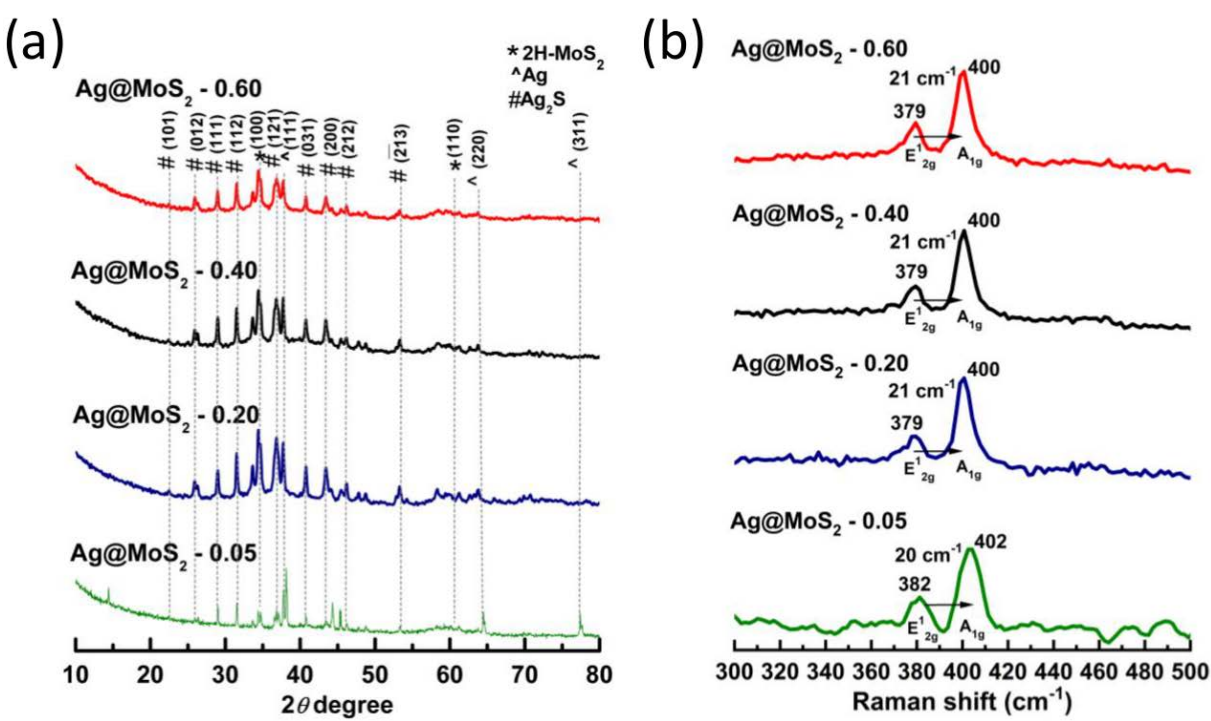

**Fig. 2.** (a) Powder XRD patterns and (b) Raman spectroscopy of the hybrids.

We used X-ray photoelectron spectroscopy (XPS) to characterize the composition of the hybrids' surface. XPS is known to penetrate a few nm into the surface, allowing for the study of the exposed surface of the nanostructures and the few atomic layers beneath (Fig. 3). The XPS detected S$^{2-}$ in both MoS$_2$ and Ag$_2$S (Figs. S3–S6). For Ag, an incremental increase in the oxidation state is associated with a negative shift in the XPS binding energies [54]. The data show the Ag 3d states (Fig. 3a) and using Auger spectroscopy (Fig. 3b) we could also resolve the fraction of metallic silver compared to Ag(I), [55,56], which is the evidence for the Ag$_2$S phase detected in the XRD patterns. Mo 3d exhibited the typical peaks at a binding energy of ~ 229 eV, which correspond to the Mo(IV) valence state in $MoS_2$ (Fig. 3c). The overall surface composition is presented in the bar graph in Fig. 3(d), showing that the sulfidation of Ag is enhanced as the ATM concentration increases.

Fig. 3(c) shows that the Mo 3d binding energy is gradually redshifted to lower values, from free-standing $MoS_2$ to hybrid core-shell structure, corresponding to charge transfer from the core to the $MoS_2$ shell [35,36]. In our theoretical work, the shelling of transition metals (Au, Ag and Cu) with $MoS_2$ was found to be thermodynamically stable, including a charge transfer from the core to the shell such that the electronic conductivity is expected to be enhanced but without inducing a phase transition to the 1T metallic state [57]. In addition to the higher number of exposed $MoS_2$ edges, the S vacancies on the basal plane of $2H-MoS_2$, the most stable adsorbed phase on the metal cores, and the strain in the $MoS_2$, tune the binding sites for HER and can significantly enhance the catalytic activity towards the HER [35,40,57].

To summarize the findings so far, during the synthesis, the Ag cores were partially sulfidized, resulting in the coexistence of metallic Ag(0) and $Ag_2S$ (with the Ag(I) ions), whose ratio depends on the loading of the ATM precursor. At low ATM concentrations, the TEM images could not detect shelling, although Raman and XPS detected $MoS_2$. It is probable that $MoS_2$ formed as small clusters and islands, while a substantial surface area was bare Ag. This picture is consistent with the ICP analysis which shows low Mo/Ag content (Table S2) and the lack of lattice strain in the $MoS_2$ phase of $Ag@MoS_2$-0.05, as revealed by the Raman data. Moreover, the shift in the XPS of the Mo 3d is less pronounced in the $Ag@MoS_2$-0.05 compared to the core-shell hybrid samples. More ATM resulted in increase in the $Ag_2S$ fraction and in the shelling with $MoS_2$, until a maximum was reached. For the sample with the highest ATM concentration, 0.60 mg $L^{-1}$, $Ag_2S$ was the main phase detected, and the $MoS_2$ fraction was even smaller than for 0.40 mg $L^{-1}$. The same is evident by the ICP-OES data, which probes the entire mass of the samples (see Table S2). This phenomenon may be attributed to an electrostatic repulsion between the $MoS_4^{2-}$ anions and the adsorbed shell over $Ag_2S$, similar to the $Cu_{2-x}S@MoS_2$ structures in our previous work [58]. When extended shelling was observed, the $MoS_2$ shell was defective and strained, potentially containing S vacancies known as catalytic sites for the HER. In addition, charge transfer occurred from the conductive core to the shell, and all of these parameters combined contribute towards making the hybrid nanostructures potentially improved catalysts for HER.

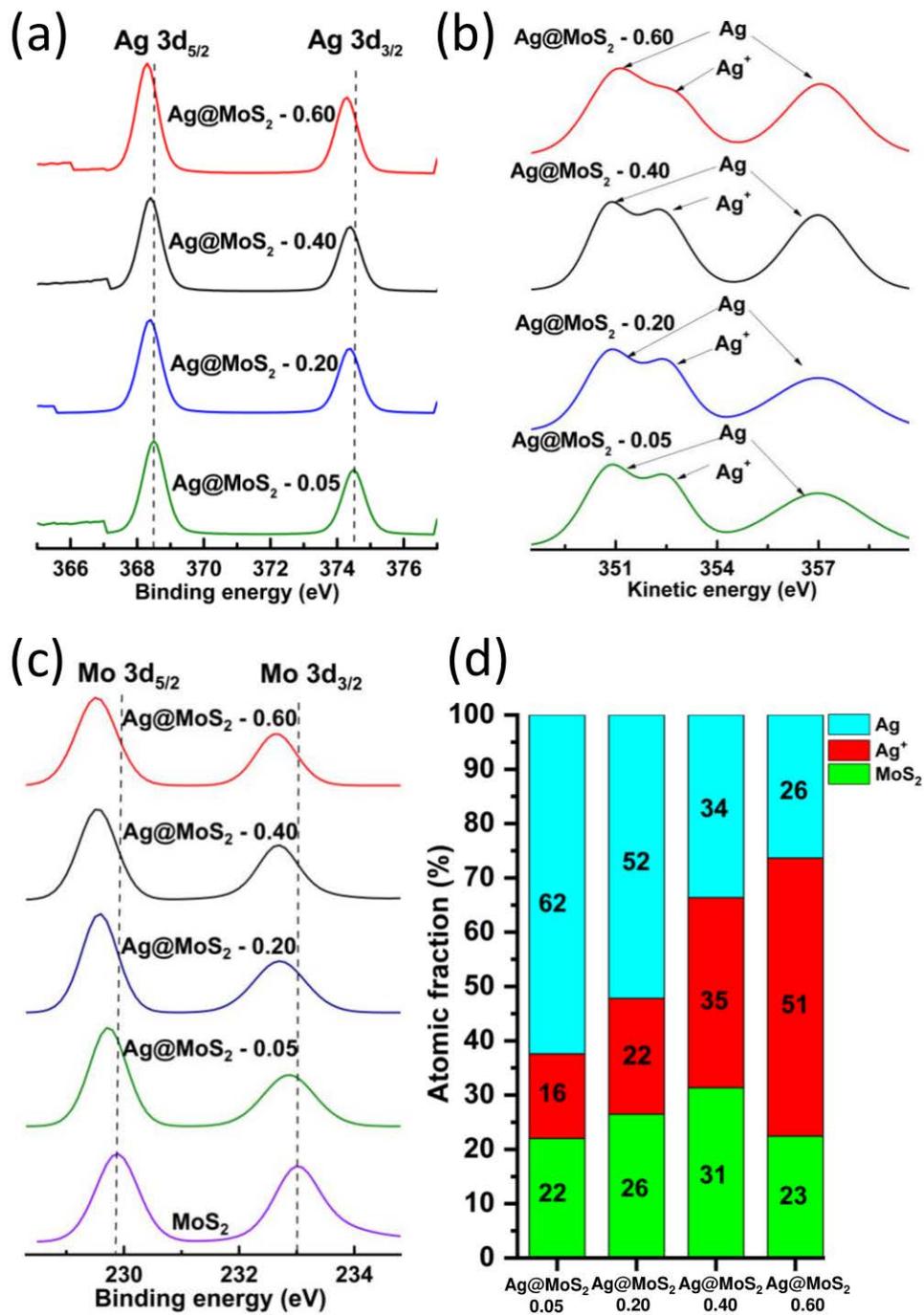

**Fig. 3.** XPS and Auger spectra of the different hybrids. (a) XPS spectrum of Ag 3d, (b) Auger spectrum of Ag MNN level, (c) XPS spectrum of Mo 3d, and (d) The chemical composition of the hybrids surface, based on the data in (a–c).

To evaluate the HER performance of the Ag@MoS$_2$ hybrids, electrochemical measurements were carried out using linear sweep voltammetry (LSV) in a standard three-electrode setup under alkaline (0.5

M KOH) and acidic (0.5 M $H_2SO_4$) media. Realizing that the HER kinetics is dramatically affected by a change of pH from acidic to alkaline, it is very beneficial to compare both environments, with the overall goal to control the HER activity by fine-tuning the active site–electrolyte interactions [12,37,38,59]. For comparison, Ag nanoparticles, $Ag_2S$ nanoparticles, free-standing $MoS_2$ and commercial 20% Pt/C were also tested (Figs. 4 and 5 and Table 1). Under alkaline conditions, the hybrid catalysts showed a significantly lower overpotential than the individual components and reduced Tafel slopes (Fig. 4a and b). Among the hybrid samples, the least-shelled hybrids, prepared with the lowest ATM concentration (0.05 mg $L^{-1}$), showed the lowest activity with a large overpotential of $\eta = 610$ mV at 10 mA $cm^{-2}$, similar to bare Ag nanoparticles ($\eta = 615$ mV). Ag@$MoS_2$-0.40 showed a significantly higher catalytic activity than the other hybrids, exhibiting an overpotential of only 260 mV at 10 mA $cm^{-2}$ and a lower Tafel slope of 120 mV $dec^{-1}$, indicating faster intrinsic catalytic kinetics. Electrochemical impedance spectroscopy (EIS) was conducted, and the Nyquist plots were fitted with an equivalent R(RC) circuit to estimate the charge transfer resistance, $R_{ct}$ (Fig. 4c and Scheme S1). The electrochemically active surface area (ECSA) was estimated by the measurement of double-layer capacitance ($C_{dl}$) in the non-faradaic potential region (Fig. 4d and Fig. S7), which is proportional to the effective active surface area, thus providing an evaluation of the degree of electrolyte-accessible electrode surface [38,40]. Both the $R_{ct}$ and the $C_{dl}$ values of Ag@$MoS_2$-0.40 were the best of all the samples in the set, indicates for its high degree of accessible active sites with faster electron transfer kinetics maintaining the trend seen in the overpotential and Tafel slope values.

**Table 1**. A summary of the electrocatalytic measurements in acidic and alkaline media

| Material | 0.5M KOH | | 0.5M $H_2SO_4$ | |
|---|---|---|---|---|
| | Overpotential ($\eta$) at 10 mA $cm^{-2}$ (mV) | Tafel slope (mV $dec^{-1}$) | Overpotential ($\eta$) at 10 mA $cm^{-2}$ (mV) | Tafel slope (mV $dec^{-1}$) |
| Ag@$MoS_2$-0.05 | 610 | 230 | 455 | 111 |
| Ag@$MoS_2$-0.20 | 340 | 175 | 265 | 105 |
| Ag@$MoS_2$-0.40 | 260 | 120 | 235 | 106 |
| Ag@$MoS_2$-0.60 | 410 | 205 | 260 | 104 |
| Free standing $MoS_2$ | 490 | 210 | 420 | 150 |
| Ag | 615 | 190 | 390 | 101 |
| $Ag_2S$ | 540 | 160 | 405 | 104 |

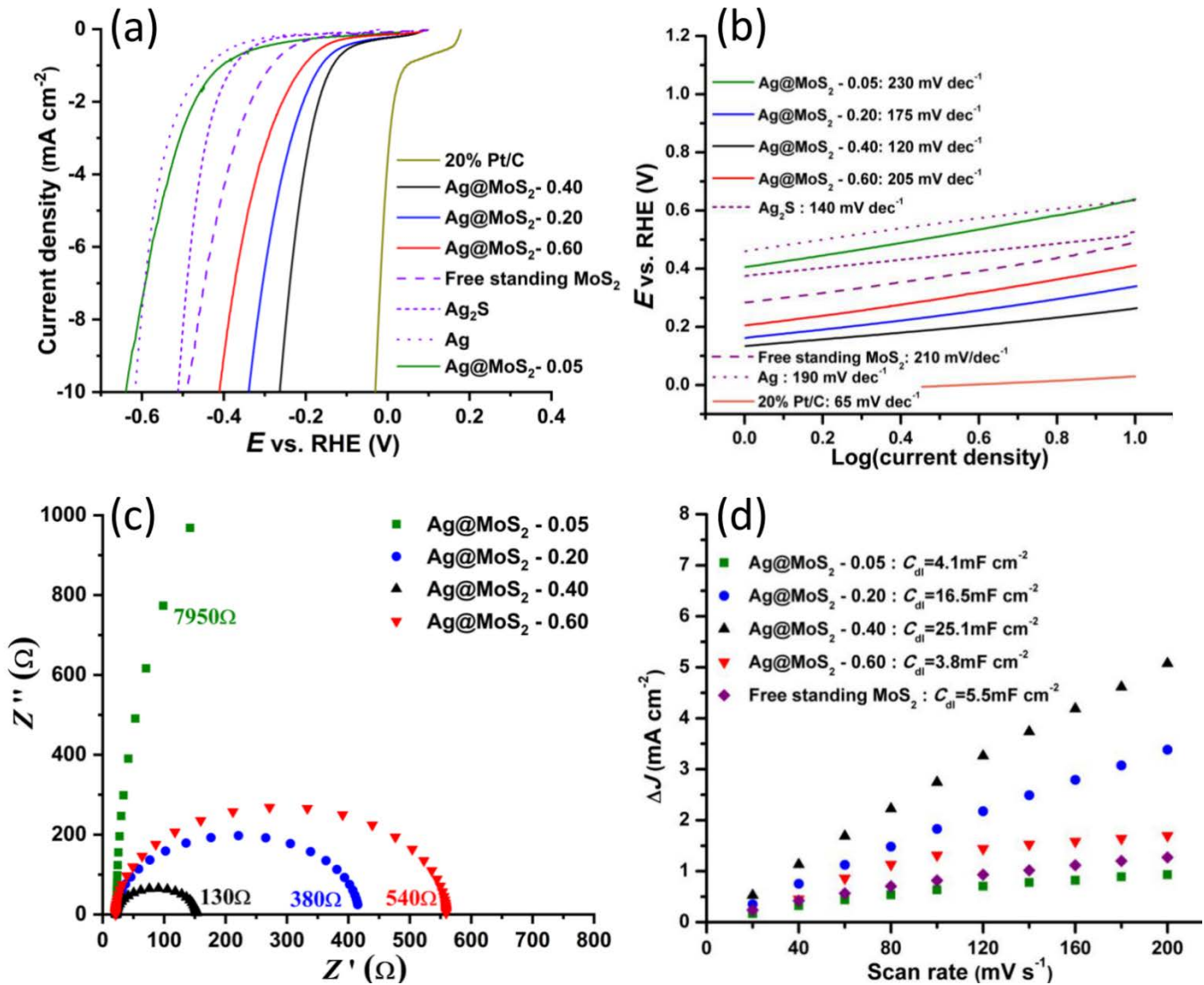

**Fig. 4**. HER performance of the various catalysts and Pt/C in 0.5M KOH. (a) LSV polarization curves acquired at a scan rate of 10 mV s$^{-1}$, (b) Corresponding Tafel slopes, (c) The Nyquist plots, (d) The double-layer capacitance as indicative of the electrochemical surface area (ECSA).

These findings indicate that Ag@MoS$_2$-0.40 represents optimal synthesis conditions for the formation of active catalysts. In alkaline HER conditions, the hydrogen intermediate is formed by the initial water dissociation step: H--OH + e$^-$ ⟶ H$_{(ads)}$ + OH, and therefore the catalytic activity is controlled by the ability of the surface to successfully dissociate the water molecules and supply protons to the catalytic reaction. In addition, it is important to avoid poisoning of the catalytic sites by OH$^-$ as well as to optimize the free energy of hydrogen binding, $\Delta G_H{}^*$, which is a key parameter in acidic conditions [6,60]. While Ag(0) hardly binds hydrogen [13], and is therefore expected to be almost inert for HER, the oxophilic Ag(I) sites may act as promoters to the water dissociation in a similar way to the recently demonstrated mechanism for Cu(I)--OH$_2$ [59]. We suggest that the hybrids exhibit a cooperative mechanism, where Ag(I) serves to decrease the energy barrier for water dissociation, H--OH, whereas the nearby MoS$_2$ sites serve to collect and recombine H$_{ads}^*$ to form H$_2$, thus facilitating the critical Volmer step. The proposed mechanism is supported by the evident improved Tafel slope and fast electron transfer for the hybrids. The right balance between the two components (Ag$_2$S and MoS$_2$) is essential, to avoid situations where the Ag(I) content is too low to overcome the transition energy barrier for water dissociation, or when the surface of the hybrid is enriched with Ag(I) but with only a few surrounding MoS$_2$ sites such that the M-H$_{add}$ formation is hindered. In addition, retaining a certain fraction of the original metallic core will be advantageous in increasing the conductivity, since both Ag$_2$S and MoS$_2$ are semiconductors [61,62]. Recent studies also demonstrated that introducing a second active component for water dissociation to a substrate with near-optimal M-H$_{ads}$ energetics can greatly improve the alkaline HER kinetics (such as Ni(OH)$_2$ decorated by Pt co-catalysts [12]). Another point is the highest $C_{dl}$ exhibited by Ag@MoS$_2$-0.40 is ~6 times higher than the barely covered hybrid Ag@MoS$_2$-0.05, which indicates a higher degree of electrolyte-accessible electrode sites to enhance the HER. Among the hybrids, Ag@MoS$_2$-0.40 exhibited $R_{ct}$ of 120 Ω, the lowest in the set, indicating its faster interfacial electron-transfer kinetics that promotes the sluggish electron-coupled water dissociation step (Volmer, H$_2$O + e$^-$ → H$_{ads}$ + OH$^-$), as implied by the lowest Tafel slope of the Ag@MoS$_2$-0.40 catalyst. The HER performance of the hybrid is better than or comparable to that of the state-of-the-art HER catalysts in alkaline solution (see the comparison in Table S4).

To further elucidate the mechanism of the catalytic enhancement, we considered the catalytic activity of the hybrids in an acidic medium. If the proposed mechanism is correct, then its main advantage is for the catalytic reaction in alkaline, where the water dissociation is a critical parameter. In contrast, under acidic conditions, the HER mechanism progresses by the following path: H$^+$ + e$^-$ → M-H* + H$^+$ +e$^-$ → H$_2$. Therefore, the proton donor is not H$_2$O, making the water dissociation irrelevant and allowing the H-adsorption ability of the catalyst sites (Δ$G_H$*) to control the HER activity (the well-known volcano plots) [63]. We predict that the ΔG$_H$* values will be close for the various hybrids, since the strained and defective MoS$_2$ layer in them is similar in properties. As shown in Fig. 5, the Ag@MoS$_2$ hybrids (Ag@MoS$_2$-0.60, Ag@MoS$_2$-0.40, and Ag@MoS$_2$-0.20) showed enhanced and similar activity relative to the sole components and to the Ag@MoS$_2$-0.05. The results support our hypothesis, that the H-adsorption properties of the hybrids surface are similar, stemming from the defective shell, while the Ag(I) mainly provide additional sites for water dissociation. It is especially interesting to note that when the shell is not strained, i.e., Ag@MoS$_2$-0.05, or when free-standing MoS$_2$ is used, the catalytic performance is low, demonstrating the importance of hybrid formation.

Electrocatalytic stability of the hybrid was also tested after continuous CV cycles in both acidic (Fig. S8) and alkaline media (Fig. S9). The results show excellent stability in acidic medium and moderate stability in alkaline. In both cases the initial morphology is maintained (Figs. S8 and S9), indicating that the durability of the modified electrode is affected by the adherence to the substrate, which can be improved using other casted substrates such as carbon cloth [64].

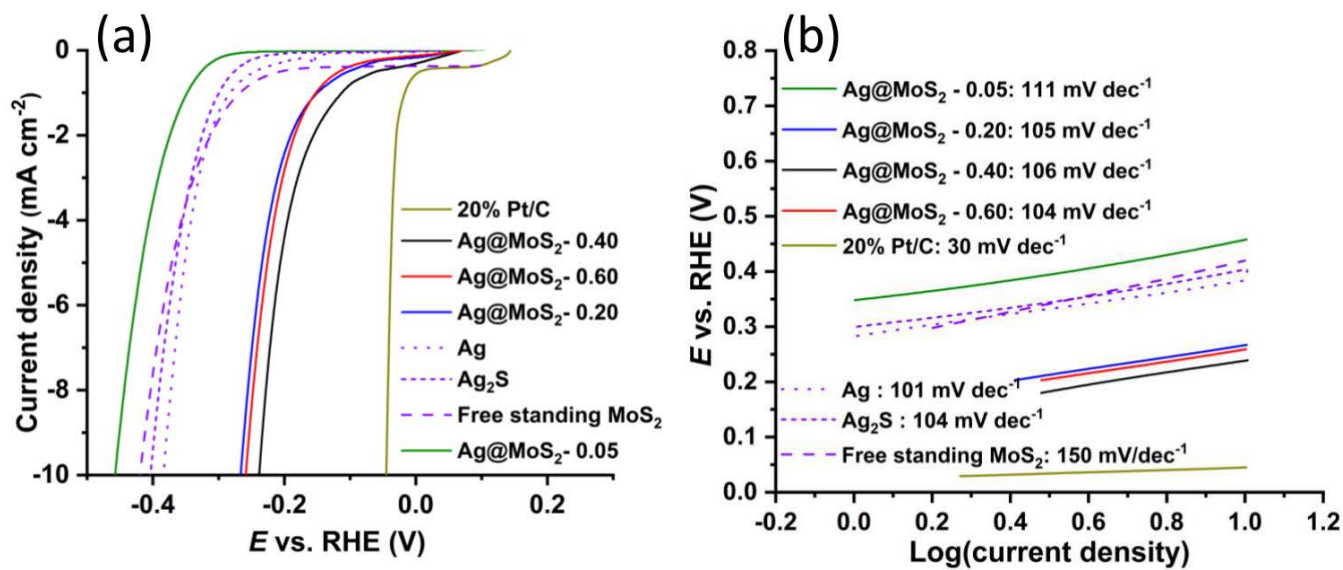

**Fig. 5.** HER performance of the various catalysts and Pt/C in 0.5M H$_2$SO$_4$. (a) LSV polarization curves acquired at a scan rate of 10 mV s$^{-1}$ (b) Corresponding Tafel slopes.

## 4. Conclusions

In summary, we found that shelling Ag core with curved MoS$_2$ layer provided efficient active sites for HER, better than the sole components. The MoS$_2$ shell produced is strained and defective, potentially comprising S vacancies known to be active catalytic sites. By varying the loading ratio of the precursors, we controlled the surface composition of the core-shell heterostructure, adding coverage of MoS$_2$ but also sulfidizing the Ag(0) core to Ag$_2$S. The hybrids exhibit significantly enhanced catalytic activity towards the hydrogen evolution activity in alkaline media, which we attribute to a cooperative effect of interfaces where water dissociation occurs at the Ag(I) sites while the recombination of H$_{ads}$ to produce H$_2$ is associated with a strained and defective MoS$_2$ layer. This work contributes to the ongoing challenge in understanding the principles for designing noble metal-free catalysts to promote the HER, particularly in water-alkaline electrolysis.

## Declaration of Competing Interest

The authors declare that they have no known competing financial interests or personal relationships that could have appeared to influence the work reported in this paper.

## Acknowledgments

This research was supported by the United States – Israel Binational Science Foundation (BSF), Jerusalem, Israel and the United States National Science Foundation (NSF) grant 2017642, and partly from the Israeli Atomic Energy Commission–Prof. A. Pazy joint foundation, ID126-2020. S.H. and R.A. acknowledge funding from the European Union's Horizon 2020 research and innovation program under the Marie Sklodowska-Curie grant agreement No 889546 as well as from the Spanish MICINN (project grant PID2019-104739GB-100/AEI/10.13039/501100011033). R.A. also support the funding from the European Union H2020 program Graphene Flagship CORE3 (881603). Some of the TEM measurements were performed in the Laboratorio de Microscopias Avanzadas (LMA) at the Universidad de Zaragoza (Spain).

**Appendix A. Supplementary data**

Supplementary data to this article can be found online